\documentclass[aps,superscriptaddress,nofootinbib,11pt]{revtex4-1} 
\pdfoutput=1

\usepackage{dcolumn}    
\usepackage{pstricks}
\usepackage{color}
\usepackage{graphicx}
\usepackage{amsmath}
\usepackage{mathtools}
\usepackage[T1]{fontenc}
\usepackage[utf8]{inputenc}
\usepackage{amssymb}
\usepackage[colorlinks=true,allcolors=black, pdfborder={0 0 0}]{hyperref}
\usepackage{siunitx}
\usepackage{mdframed}

\usepackage{booktabs}

\allowdisplaybreaks

\def\MGUT{M_\mathrm{GUT}}

\def\MZ{M_Z}
\def\SO10{\text{SO}(10)}

\def\SU{\,\text{SU}}

\newcommand{\rep}[1]{\mathbf{#1}}
\newcommand{\repb}[1]{\mathbf{\overline{#1}}}
\newcommand{\Lag}[1]{\mathcal{L}_\text{#1}}

\newcommand{\ii}{\mathrm{i}}

\definecolor{darkred}{rgb}{0.6,0,0}

\begin{document}

\title{Flavor Symmetries in the Yukawa Sector of Non-Supersymmetric SO(10):\\ Numerical Fits Using Renormalization Group Running}	
	
\author{Tommy Ohlsson}
\email{tohlsson@kth.se}
\affiliation{Department of Physics,
	School of Engineering Sciences,
	KTH Royal Institute of Technology,
	AlbaNova University Center,
	Roslagstullsbacken 21,
	SE--106 91 Stockholm,
	Sweden}
\affiliation{The Oskar Klein Centre for Cosmoparticle Physics,
	AlbaNova University Center,
	Roslagstullsbacken 21,
	SE--106 91 Stockholm,
	Sweden}
\affiliation{University of Iceland, 
	Science Institute, 
	Dunhaga 3, 
	IS--107 Reykjavik, 
	Iceland}

\author{Marcus Pernow}
\email{pernow@kth.se}
\affiliation{Department of Physics,
	School of Engineering Sciences,
	KTH Royal Institute of Technology,
	AlbaNova University Center,
	Roslagstullsbacken 21,
	SE--106 91 Stockholm,
	Sweden}
\affiliation{The Oskar Klein Centre for Cosmoparticle Physics,
	AlbaNova University Center,
	Roslagstullsbacken 21,
	SE--106 91 Stockholm,
	Sweden}

\begin{abstract}
We consider a class of $\SO10$ models with flavor symmetries in the Yukawa sector and investigate their viability by performing numerical fits to the fermion masses and mixing parameters. The fitting procedure involves a top-down approach in which we solve the renormalization group equations from the scale of grand unification down to the electroweak scale. This allows the intermediate scale right-handed neutrinos and scalar triplet, involved in the type I and II seesaw mechanisms, to be integrated out at their corresponding mass scales, leading to a correct renormalization group running. The result is that, of the 14 models considered, only two are able to fit the known data well. Both these two models correspond to $\mathbb{Z}_2$ symmetries. In addition to being able to fit the fermion masses and mixing parameters, they provide predictions for the sum of light neutrino masses and the effective neutrinoless double beta decay mass parameter, which are both within current observational bounds. 
\end{abstract}

\maketitle

\section{Introduction}
One way of extending the Standard Model (SM) is to assume that, at some high energy scale, a Grand Unified Theory (GUT)~\cite{Georgi:1974sy} provides a more fundamental description of particle physics. Particularly, models based on the gauge group $\SO10$~\cite{Fritzsch:1974nn} are successful in accommodating all SM fermions in one representation per generation and providing an origin for several of the features in the SM. It is also a useful class of models in which to explore the outstanding questions of the SM, such as neutrino masses through the type I and II seesaw mechanisms~\cite{Minkowski:1977sc,GellMann:1980vs,Mohapatra:1979ia,Yanagida:1979,Schechter:1980gr,Magg:1980ut,Lazarides:1980nt,Mohapatra:1980yp}.

The Yukawa sector of $\SO10$ models with the type I or II seesaw mechanisms has been explored in numerical fits, both in the supersymmetric case~\cite{Bertolini:2004eq,Babu:2005ia,Bertolini:2006pe,Bajc:2008dc,Altarelli:2010at,Dueck:2013gca,Fukuyama:2015kra,Ferreira:2015jpa,Deppisch:2018flu} and the non-supersymmetric one~\cite{Joshipura:2011nn,Altarelli:2013aqa,Dueck:2013gca,Meloni:2014rga,Babu:2015bna,Babu:2016bmy,Meloni:2016rnt,Ohlsson:2018qpt,Boucenna:2018wjc,Ohlsson:2019sja}. For a more complete list of works, see Ref.~\cite{Deppisch:2018flu} and references therein. The result in both the supersymmetric and non-supersymmetric cases is that, depending on the scalar content in the Yukawa sector, it is in general possible to accommodate the known masses and mixing parameters of the SM in an $\SO10$ model. For the neutrino masses, the best fit is achieved with a combination of the type I and II seesaw mechanisms, which naturally occurs in $\SO10$.

Another outstanding question in the SM is the origin of the observed masses and mixing parameters. This has been explored extensively in the framework of $\SO10$ models. One way of motivating the observed patterns is to impose an additional flavor symmetry, also known as a horizontal symmetry, which constrains the Yukawa matrices and thereby imposes a flavor pattern~\cite{Papageorgiu:1994qw,Albright:1995yp,Chen:2000fp,Chen:2001pra,Chen:2002pa,King:2003rf,King:2009mk,Dutta:2009bj,Lam:2014kga,Ivanov:2015xss,Ferreira:2015jpa,Bajc:2016eiw,Meloni:2017cig}. Other models with zero Yukawa matrix textures are also discussed in Refs.~\cite{Ramond:1993kv,Dutta:2004zh,Fukuyama:2007ri}. Additional attempts to explain the observed data involve introducing additional fermion fields~\cite{Malinsky:2007qy,Heinze:2010du,Babu:2016cri} and using the clockwork mechanism~\cite{Babu:2020tnf}.

In Ref.~\cite{Ferreira:2015jpa}, the authors considered all possible flavor symmetries based on phase transformations in the Yukawa sector of a supersymmetric $\SO10$-based model with the scalar representations $\rep{10}_\text{H}$, $\repb{126}_\text{H}$, and $\rep{120}_\text{H}$. This led to 14 different models with differing Yukawa matrix textures. They performed numerical fits to these models at the GUT scale $\MGUT$ and found that only two models allow for realistic fermion masses and mixing parameters. In this work, we perform similar fits to these models, but in the absence of supersymmetry. Furthermore, and more importantly, instead of evolving the data to $\MGUT$ and fitting the model parameters to those, we solve the renormalization group equations (RGEs) from $\MGUT$ down to the electroweak scale $\MZ$ for each sampled point and fit at $\MZ$. This approach immensely increases the complexity of the problem and hence the computational time required to perform the fits, but allows for appropriately integrating out heavy fields with masses between $\MZ$ and $\MGUT$ and matching the parameters of the theories above and below that scale. Not taking this into account is a commonly used approximation, which nevertheless can significantly alter the results~\cite{Dueck:2013gca}.

This work is structured as follows. In Sec.~\ref{sec:symms_models}, we briefly review the models and discuss their parametrization. Then, in Sec.~\ref{sec:procedure}, we describe the fitting procedure, and in Sec.~\ref{sec:results}, we present the results of the fits. Finally, in Sec.~\ref{sec:conclusion}, we summarize and conclude. A full list of the 14 models is presented in App.~\ref{sec:app_models}.

\section{Flavor Symmetries and Models}\label{sec:symms_models}
\subsection{Models}\label{sec:models}
In this work, we consider a setup in which the $\SO10$ Yukawa sector contains scalar fields belonging to the $\rep{10}_\text{H}$, $\repb{126}_\text{H}$, and $\rep{120}_\text{H}$ representations. As such, the Lagrangian is given by
\begin{equation}
\Lag{Yuk} = -\rep{16}_\text{F}( Y_{10} \rep{10}_\text{H}  + Y_{126} \repb{126}_\text{H}+ Y_{120} \rep{120}_\text{H}) \rep{16}_\text{F} + \text{h.c.},
\end{equation}
where the fermions are placed in the $\rep{16}_\text{F}$ representations and $Y_i$ are $3\times 3$ matrices of Yukawa couplings in flavor space. Due to the structure of the $\SO10$ coupling, the Yukawa matrices are such that $Y_{10}$ and $Y_{126}$ are symmetric, while $Y_{120}$ is antisymmetric. The different models that we investigate correspond to different textures of zeros of these three matrices. These textures originate in symmetries under phase transformations of the three fermion generations and the scalars in the Yukawa sector. In this framework, the allowed symmetries are $\mathbb{Z}_2$, $\mathbb{Z}_3$, $\mathbb{Z}_4$, $\text{U}(1)$, and $\mathbb{Z}_2\times\mathbb{Z}_2$. The complete derivation and discussion of these symmetries are presented in Refs.~\cite{Ivanov:2015xss, Ferreira:2015jpa}. Here, we reiterate the main points. 

The symmetries are determined by the possible rephasings of each of the three generations independently, as well the three scalar multiplets. Any specific Yukawa matrix texture will impose constraints on these phases. The remaining free phases determine the symmetry group that a texture allows for. Further constraints on the Yukawa matrices are that $Y_{126}$ may not be singular (since we need to invert it for type I seesaw), all three Yukawa matrices are non-zero (since otherwise we would not have a model with all three scalar representations), and no generation decouples in a linear combination of Yukawa matrices (since we would then not be able to accommodate quark mixing). Furthermore, two models that differ by a redefinition of the labeling of generations are considered equivalent. With these constraints, there are 14 possible inequivalent models, which we list in App.~\ref{sec:app_models} and summarize in Tab.~\ref{tab:models} using the same naming convention as in Ref.~\cite{Ferreira:2015jpa}. As an example, one of the models (denoted model~A) is given by

\begin{equation}
\text{A}:\quad Y_{10} \sim \begin{pmatrix}
\times & \times & 0 \\
\times & \times & 0 \\
0 & 0 & \times
\end{pmatrix}, \quad
Y_{126} \sim \begin{pmatrix}
\times & \times & 0 \\
\times & \times & 0 \\
0 & 0 & \times
\end{pmatrix},\quad 
Y_{120}\sim \begin{pmatrix}
0 & 0 & \times \\
0 & 0 & \times \\
\times & \times & 0
\end{pmatrix},
\end{equation}
where the symbol ``$\times$'' denotes a general non-zero matrix element.

The more non-zero matrix elements that a model has, the more constrained the symmetry is. Similarly, many zeroes imply a large degree of freedom in the symmetry transformations. Some of the models can be derived from other models by setting some of the elements to zero. One can, for example, derive model~A$_1$ from A in this way. All such submodels are included only if they give rise to different symmetry groups. For example, setting the $(1,2)$ element of $Y_{10}$ to zero in model A does not affect the symmetry, so that is not considered to be a separate model.

\begin{table}
\setlength{\tabcolsep}{1.5em}
    \begin{center}
    \begin{tabular}{ c c c }
    \hline\hline
         Model & Symmetry& Number of parameters\\\hline 
         A & $\mathbb{Z}_2$ & 29 \\
         A$_1$ & $\mathbb{Z}_4$ & 22 \\  
         A$_1'$ & $\text{U}(1)$ & 21 \\  
         A$_1''$ & $\text{U}(1)$ & 21 \\  
         A$_2$ & $\text{U}(1)$ & 24 \\  
         B & $\mathbb{Z}_2$ & 25 \\  
         C & $\mathbb{Z}_2$ & 23 \\   
         D$_1$ & $\mathbb{Z}_3$ & 24 \\  
         D$_2$ & $\mathbb{Z}_3$ & 24 \\  
         D$_3$ & $\mathbb{Z}_3$ & 24 \\  
         D$_1'$ & $\text{U}(1)$ & 22 \\  
         D$_2'$ & $\text{U}(1)$ & 22 \\  
         D$_3'$ & $\text{U}(1)$ & 22 \\  
         E & $\mathbb{Z}_2\times\mathbb{Z}_2$ & 21 \\  \hline\hline
    \end{tabular}
    \end{center}
    \caption{\label{tab:models}The models listed in App.~\ref{sec:app_models} and their associated symmetry groups and numbers of parameters. For details about the parametrization, see Sec.~\ref{sec:params}.}
\end{table}

We assume a unification scale of $\MGUT=10^{16}\,\text{GeV}$, at which $\SO10$ is assumed to spontaneously break to the SM. As in Refs.~\cite{Dueck:2013gca,Ohlsson:2019sja}, we do not impose gauge coupling unification, but assume that this is taken care of by some other new physics at an intermediate scale~\cite{Kynshi:1993zr,Ma:2008cu,Frigerio:2009wf,Kadastik:2009cu,Parida:2011wh,Aizawa:2014iea,Hagedorn:2016dze,Parida:2016hln,Boucenna:2018wjc,Banerjee:2020ule}. Although this will, in general, affect the RG running of the fermion observables through the contributions of the gauge couplings to the RGEs, we assume that this change is negligible to the ability to fit the models. This assumption is based on the fact that varying the unification scale by an order of magnitude has a small effect on the goodness of fit (see Fig.~3 of Ref.~\cite{Ohlsson:2019sja}), while the neutrino thresholds (\textit{i.e.}~the integrating out of the heavy right-handed neutrinos or scalar triplet) have a large direct effect, as shown in Tab.~\ref{tab:results} of Sec.~\ref{sec:results}, since the RG running of top-down and bottom-up are obviously very different. The changes of the RG running due to the new physics that achieves gauge coupling unification is highly dependent on the choice of \textit{e.g.} intermediate-scale scalar fields and we do not wish to specify any specific model.

\subsection{Parametrization}\label{sec:params}
From the $\SO10$ Yukawa matrices, we need to transform the parameters to the SM Yukawa matrices. To that end, we use the fact that the vevs of the $\SU(2)_\text{L}$ doublets contained within the three scalar representations make up the vev of the SM Higgs $v_\text{SM}$. In order to have two separate vevs from the $\rep{10}_\text{H}$ without requiring an additional Yukawa coupling, we complexify the $\rep{10}_\text{H}$ and impose a Peccei--Quinn (PQ) symmetry~\cite{Bajc:2005zf}. The matching conditions  are then given by~\cite{Altarelli:2013aqa,Dueck:2013gca,Joshipura:2011nn,Babu:2015bna}
\begin{align}
v_\text{SM}Y_u &= v^u_{10} Y_{10} + v^u_{126} Y_{126} + (v^u_{120_1} + v^u_{120_2})Y_{120}, \\
v_\text{SM}Y_d &= v^d_{10} Y_{10} + v^d_{126} Y_{126} + (v^d_{120_1} + v^d_{120_2})Y_{120}, \\
v_\text{SM}Y_D &= v^u_{10} Y_{10} -3 v^u_{126} Y_{126} + (v^u_{120_1} - 3v^u_{120_2})Y_{120}, \\
v_\text{SM}Y_\ell &= v^d_{10} Y_{10} -3 v^d_{126} Y_{126} + (v^d_{120_1} - 3v^d_{120_2})Y_{120}, \\
M_R &= v_R Y_{126}, \\
M_L &= v_L Y_{126},
\end{align}
where $Y_u$, $Y_d$, $Y_D$, and $Y_\ell$ are the Yukawa coupling matrices of the up-type quarks, down-type quarks, neutrinos, and charged leptons, respectively. In addition, $M_R$ is the mass matrix of the right-handed neutrinos and $M_L$ is the contribution from type II seesaw to the light neutrino mass matrix. The vevs $v^{u,d}_{10,126,120}$ are the vevs of the $\SU(2)_\text{L}$ doublets contained in the $\rep{10}_\text{H}$, $\repb{126}_\text{H}$, and $\rep{120}_\text{H}$, while $v_R$ is the vev of the SM singlet in $\repb{126}_H$ that gives mass to the heavy right-handed neutrinos and $v_L$ is the vev of the $\SU(2)_\text{L}$ triplet involved in the type II seesaw contribution to the neutrino masses.

For convenience, we reparametrize using $H=Y_{10}v^d_{10}/v_\text{SM}$, $F = Y_{126}v^d_{126}/v_\text{SM}$, and $G=Y_{120}(v^d_{120_1}+v^d_{120_2})/v_\text{SM}$ and write the Yukawa and neutrino mass matrices as
\begin{align}
Y_u &= r(H+sF+t_uG),\label{eq:mass_matrices1} \\
Y_d &= H + F + G,\\
Y_D &= r(H-3sF+t_D G),\\
Y_\ell &= H - 3F + t_\ell G,\\
M_R &= r_R F, \\
M_L &= r_L F.
\end{align}
Since the matrices $H$, $F$, and $G$ are proportional to the Yukawa matrices $Y_{10}$, $Y_{126}$, and $Y_{120}$, respectively, they have, of course, the same textures.

The parameters other than the Yukawa matrix elements that are to be sampled are then $r$, $s$, $t_u$, $t_D$, $t_\ell$, $r_R$, and $r_L$. Of these, we assume $r$ and $r_R$ to be real, while the others are complex, giving a total of twelve parameters related to the vevs. Furthermore, we need $v^d_{126}$ and the mass $M_\Delta$ of the $\SU(2)_\text{L}$ triplet for the matching to the effective neutrino mass matrix, thus adding two parameters. For more details on the effective neutrino mass matrix, see Sec.~\ref{sec:procedure}.

To count the number of parameters in the matrices $H$, $F$, and $G$, we first note that they, in general, have complex matrix elements and that the (anti)symmetry of them restrict the number of free parameters. One can, without loss of generality, transform one of $H$ or $F$ to a real and diagonal matrix. This same transformation must also be applied to the other two matrices, so it should only be performed if it decreases the total number of parameters in the model. 

Since the vevs of the $\SU(2)_\text{L}$ multiplets all contribute to the electroweak gauge boson masses after electroweak symmetry breaking, they need to add up in quadrature to the SM Higgs vev. The parametrization used leaves some freedom in this constraint, since not all vevs are sampled. Further imposing a perturbativity bound of $4\pi$ on the Yukawa couplings, we have the inequality
\begin{equation}\label{eq:vev_ineq}
\begin{split}
& \max|H|^2(1 + r^2) +  |v^d_{126}|^2 \left(\frac{4\pi}{v_\text{SM}}\right)^2 (1+r^2|s|^2 + 2|r_L|^2) \\
&+ \max |G|^2 \frac{r^2}{16} \left(|t_D + 3t_u|^2 + |t_u - t_D|^2\right) \leq 16\pi^2,
\end{split}
\end{equation}
where $\max|H|^2$ denotes the largest absolute value of the elements of $H$ and similarly for $G$. The factor of $2$ in front of $|r_L|^2$ is due to how the triplet couples to the gauge bosons. This inequality needs to be checked after each fit to make sure that the found parameter values are acceptable.

Furthermore, there are experimental bounds on the parameters related to the $\SU(2)_\text{L}$ triplet: $v_L\lesssim 1~\text{GeV}$ and $M_\Delta\gtrsim 1~\text{TeV}$~\cite{Perez:2008ha,Ferreira:2019qpf}. Finally, the vev $v_R$ should be set to $\MGUT$. However, we allow this parameter to vary by around an order of magnitude from $\MGUT$, which can be motivated by the existence of threshold corrections and higher-dimensional operators~\cite{Weinberg:1980wa,Hall:1980kf,Ellis:1979fg,Hill:1983xh,Shafi:1983gz}.

\section{Fitting Procedure}\label{sec:procedure}
When fitting the models to data, we use the known values of the fermion masses and mixing parameters at the electroweak scale $\MZ$ and perform top-down RG running for each sampled set of parameters. This is in contrast to the procedure used in Ref.~\cite{Ferreira:2015jpa}, where the fits were performed with values that had been evolved to $\MGUT$ using bottom-up RG running prior to fitting. Our approach of solving the RGEs for the sampled parameters has the benefit of correctly taking into account the mass thresholds of the heavy right-handed neutrinos and the scalar triplet for type I and II seesaw mechanisms, respectively. That is not possible when solving the RGEs before the fitting procedure, since the masses of these heavy fields are sampled and are thus unknown prior to the sampling process. However, the approach used in this work significantly increases the computational resources required, since the RGEs are solved for every sampled point in the parameter space. The computational time required is increased by a factor of about 1000 compared to the approximation of extrapolating the data to $\MGUT$ before fitting.

For each sampled point in the parameter space of the $\SO10$ Yukawa matrices and vevs, we transform the parameters to the SM Yukawa matrices and solve the RGEs at one-loop order~\cite{Jones:1981we,Machacek:1983tz,Machacek:1983fi,Machacek:1984zw,Antusch:2002rr,Antusch:2005gp,Chao:2006ye,Schmidt:2007nq,Dueck:2013gca} down to $\MZ$. When the RGE solver encounters a threshold corresponding to a heavy right-handed neutrino or the scalar triplet (involved in the type I and II seesaw mechanisms, respectively), the corresponding heavy field is integrated out and all relevant parameters are transformed to the effective field theory, before the RGE solver proceeds. A more detailed description of this procedure may be found in Ref.~\cite{Ohlsson:2019sja}, on which our code is based.

The effective neutrino mass matrix $\kappa$ is formed by starting with a matrix of zeroes. At an energy scale corresponding to the mass $M_i$ of a right-handed neutrino, it is updated according to~\cite{Antusch:2002rr,Antusch:2005gp}
\begin{equation}\label{eq:tI}
\kappa \rightarrow \kappa + \frac{2}{M_i} \left(Y_D^{(i)}\right)^T \left(Y_D^{(i)}\right),
\end{equation}
where $Y_D^{(i)}$ is the the $i$th row of $Y_D$ in a basis in which $M_R$ is diagonal. In this basis, the $i$th row of $Y_D$ is removed as well as the $i$th column and row of $M_R$. At the energy scale of the scalar triplet mass, $\kappa$ is similarly updated according to~\cite{Schmidt:2007nq}
\begin{equation}\label{eq:tII}
\kappa \rightarrow \kappa - 4 \frac{v_L}{v_\text{SM}^2}Y_\Delta
\end{equation}
and $Y_\Delta$ is then removed from the parameters in the RG running. 

The data that the parameters are fitted to are displayed in Tab.~\ref{tab:data}. All quark and charged lepton masses as well as the Higgs quartic coupling are adopted from Ref.~\cite{Huang:2020hdv}. Neutrino masses and leptonic mixing angles are the normal neutrino mass ordering\footnote{We do not consider inverted neutrino mass ordering, since previous similar fits~\cite{Ohlsson:2018qpt} have shown that only normal ordering results in acceptable fits.} values from the July 2020 version of NuFIT~\cite{Esteban:2020cvm}. Finally, CKM parameters are taken from the Summer 2019 version of CKMFitter~\cite{Charles:2004jd}. Some observables (\textit{e.g.} the charged lepton masses) are known to an extremely high precision, which makes the fit practically very difficult. To remedy this, we have artificially enlarged the errors to be at least 5~\% of the central values. This applies to all observables except for $m_u$, $m_d$, and $m_s$.

\begin{table}
\setlength{\tabcolsep}{1.5em}
    \begin{center}
    \begin{tabular}{ c @{}S@{} @{}S@{} }
    \hline \hline
         Observable & {Central value} & {Error}\\\hline 
         $m_u$ [MeV] & 1.23 & 0.21 \\
         $m_c$ [GeV] & 0.620 & 0.31* \\
         $m_t$ [GeV] & 168 & 8.4* \\
         $m_d$ [MeV] & 2.67 & 0.19 \\
         $m_s$ [MeV] & 53.2 & 4.6 \\
         $m_b$ [GeV] & 2.84 & 0.14* \\
         $m_e$ [MeV] & 0.483 & 0.024* \\
         $m_\mu$ [GeV] & 0.102 & 0.051* \\
         $m_\tau$ [GeV] & 1.73 & 0.086* \\
         $\Delta m^2_{21}$ [$10^{-5}\ \text{eV}^2$] & 7.42 & 0.37* \\
         $\Delta m^2_{31}$ [$10^{-3}\ \text{eV}^2$] & 2.51 & 0.13* \\
         $|V_{12}|$ & 0.225 & 0.011* \\
         $|V_{13}|$ & 0.00368 & 0.00018* \\
         $|V_{23}|$ & 0.0416 & 0.0021* \\
         $J\,[10^{-5}]$ & 3.06 & 0.15* \\
         $\sin^2\theta^\ell_{12}$ & 0.304 & 0.015* \\
         $\sin^2\theta^\ell_{13}$ & 0.0221 & 0.0011* \\
         $\sin^2\theta^\ell_{23}$ & 0.570 & 0.028* \\
         $\lambda$ & 0.558 & 0.028*\\ \hline\hline
    \end{tabular}
    \end{center}
    \caption{\label{tab:data}Data for the SM observables at $\MZ$ and their corresponding errors used in the fits. Quark masses, charged lepton masses, and Higgs quartic coupling are adopted from Ref.~\cite{Huang:2020hdv}, neutrino masses and leptonic mixing angles from the July 2020 version of NuFIT~\cite{Esteban:2020cvm}, and CKM parameters from the Summer 2019 version of CKMFitter~\cite{Charles:2004jd}. An asterisk denotes an error that has been enlarged to 5~\% of the corresponding central value.}
\end{table} 

After performing the RG running down to $\MZ$ and computing the values of the 19 SM observables, we form the $\chi^2$ function defined as
\begin{equation}
\chi^2 = \sum_{i=1}^{19} \left(\frac{X_i - \bar{x_i}}{\sigma_i} \right)^2,
\end{equation}
where $X_i$ is the prediction from the sampled point and $\bar{x}_i$ and $\sigma_i$ are the central value and error, respectively, as given in Tab.~\ref{tab:data}. The fitting procedure involves minimizing the $\chi^2$ function using the \texttt{Diver} code from the \texttt{ScannerBit} package~\cite{Workgroup:2017htr}, which implements a differential evolution algorithm and can be parallelized on a cluster utilizing several CPU cores. Although one can never guarantee a global minimum to be been found, we increase our confidence in the result by running the \texttt{Diver} code twice. More than this was not feasible due to the constraint on the number of core-hours available for computations. After convergence, we look for further improvements of the candidate best-fit point using a Nelder--Mead simplex algorithm~\cite{Press:1992zz}. 

For comparison, we also perform a fit to data at $\MGUT$ without the procedure for RG running of the sampled parameters discussed above. To find the data at $\MGUT$, we use the code \texttt{SMDR}~\cite{Martin:2019lqd} to perform bottom-up RG running of the data presented in Ref.~\cite{Huang:2020hdv}. The errors were computed by sampling from a Gaussian distribution with the given standard deviation, performing the RG running of all those parameters, and finally extracting the errors from the resulting distribution at $\MGUT$. Due to the fact that one cannot perform matching at right-handed neutrino and scalar triplet mass thresholds, the neutrino parameters were ignored in this procedure, and the low-energy ones were used.

During the sampling process, the elements of the matrices $H$, $F$, and $G$ are sampled logarithmically in the range $[10^{-8},10^{-1}]$, with complex phases in the interval $[0,2\pi)$. Real parameters are allowed to be positive or negative. The vev ratios $t_u$, $t_\ell$, and $t_D$ are sampled uniformly in $[0,100]$ with complex phases in $[0,2\pi)$. We sample $r$ uniformly in $[-150,150]$ and $s$ in $[0,10]$ with a complex phase in $[0,2\pi)$. The parameter $r_R$ is sampled logarithmically in $[10^{14},10^{17}]\,\text{GeV}$, $v_d$ is sampled in $[0, 10]\,\text{GeV}$, $r_L$ is sampled logarithmically in $[10^{-1},10^{-9}]$ with a complex phase in $[0,2\pi)$, and $M_\Delta$ is sampled logarithmically in $[10^6,10^{15}]\,\text{GeV}$. These ranges are decided upon partly from knowledge on previous fits and partly to fulfill the constraints that exist. Finally, we also need the Higgs quartic coupling $\lambda$ at $\MGUT$. Due to the form of the RGEs and from knowledge on previous fits, it needs to be very close to zero at $\MGUT$. Hence, we set it as such and, in order to make sure that it stays positive for the sake of vacuum stability~\cite{Buttazzo:2013uya}, we perturb it a little after the fit without any noticeable effect on the results.

\section{Results}\label{sec:results}

The results of the fitting procedure for all models are shown in Tab.~\ref{tab:results}. In this table, we list the best $\chi^2$ values from the fits with top-down RG running as well as the best $\chi^2$ values from the fits with bottom-up RG running. Only models A and B are able to fit the data in Tab.~\ref{tab:data}, with $\chi^2$ values of $3.17$ and $25.8$, respectively. To judge how acceptable a fit is, we compare the $\chi^2$ value to the number of observables that are fitted to, which is 19 in our case.\footnote{One would expect to achieve a perfect fit when the number of parameters is the same or larger than the number of observables. However, due to the non-linearity of the investigated models, this is not correct and the statistical interpretation of $\chi^2$ per degrees of freedom does not apply~\cite{Bjorkeroth:2017ybg, Deppisch:2018flu}. Nevertheless, a $\chi^2$ value can still be used as a measure of the goodness of a fit, but instead of comparing it to the number of degrees of freedom, we compare it to the number of observables.} If they are comparable, it means that the average deviation of the predicted value from the true value is around one standard deviation. For model~A, the $\chi^2$ per number of observables is $3.17/19\simeq 0.167$, which is very good. For model~B, it is $25.8/19\simeq 1.36$, which we still consider to be acceptable, especially in comparison to the other models. The next best fits after models~A and B are found for models~$\text{A}_1$ and $\text{D}_3$, see Tab.~\ref{tab:results}. It is not surprising that models A and B would best be able to accommodate the data, since their Yukawa textures are the least restrictive of all models investigated. This is in agreement with the results of Ref.~\cite{Ferreira:2015jpa}, although those results were achieved with supersymmetry and with bottom-up RG running. Both models have a $\mathbb{Z}_2$ symmetry, which is the smallest non-trivial discrete group. 

\begin{table}
\setlength{\tabcolsep}{1.5em}
    \begin{center}
    \begin{tabular}{ c c c}
    		\hline \hline
         Model & $\chi^2$ (top-down) & $\chi^2$ (bottom-up)\\ \hline 
         $\text{A}$ & 3.17 &  $1.8\times10^{-11}$    \\
         $\text{A}_1$ & 99.7  & 271  \\
         $\text{A}_1'$ & 412  & 488 \\
         $\text{A}_1''$ & 607  & 500 \\
         $\text{A}_2$ & 213*  & 67.4 \\
         $\text{B}$ & 25.8 & $2.1\times10^{-8}$  \\
         $\text{C}$ & 151 & 112 \\
         $\text{D}_1$ & 335 & 259 \\ 
         $\text{D}_1'$ &1890  & 723 \\
         $\text{D}_2$ & 444 & 389 \\ 
         $\text{D}_2'$ & 334  & 1943  \\
         $\text{D}_3$ & 92 & 825  \\
         $\text{D}_3'$ & 2009 & 1143  \\
         $\text{E}$ & 261 &  457 \\ \hline\hline
    \end{tabular}
    \end{center}
    \caption{\label{tab:results}$\chi^2$ values for the various models, both with top-down RG running and with the approximation of fitting at $\MGUT$ (labeled as ``$\chi^2$ (bottom-up)''). An asterisk denotes a point that does not satisfy the inequality in Eq.~\eqref{eq:vev_ineq} and thus is not an acceptable point.}
\end{table}

Some of the results have parameter values that violate the inequality in Eq.~\eqref{eq:vev_ineq}, denoted by an asterisk. These are therefore not acceptable points in parameter space. Notably, model~$\text{A}_2$ yields no valid best-fit point and the best one for model~$\text{D}_1$ violates the inequality.

Since the fits with bottom-up RG running (\textit{i.e.}~using the approximation of fitting at $\MGUT$) are comparatively computationally simple, we find several best fits, but we only present one for each model, for comparison. One might worry that, if the case with top-down RG running produces worse fits than bottom-up, the larger $\chi^2$ values are only a consequence of the difficulty of the algorithm to explore the parameter space and not of the actual viability of the model. However, the fact that some of the fits (\textit{i.e.}~for the fits of models $\text{A}_1$, $\text{A}_1'$, $\text{D}_2'$, $\text{D}_3$, and $\text{E}$) are actually better with top-down RG running than bottom-up increases our confidence that they reflect the viability of the models, as does the fact that the same two models are found to fit the data well both with top-down and bottom-up RG running. The bottom-up RG running is an approximation to the top-down RG running for which the right-handed neutrino thresholds are ignored. The two RG running procedures are therefore not inverses of each other, which means that one should not expect them to produce the same result. For this reason, inserting a best-fit point from the bottom-up RG running into the top-down RG running typically produces a bad result. The fact that the best-fit points from the two procedures are usually very different shows the effect of the right-handed neutrino thresholds on the RG running.

For the two best fits of models A and B, we list the corresponding predicted values of the observables and the pulls in Tab.~\ref{tab:pulls}. Notably, both models are able to fit the quark and lepton mixing parameters well, while the largest tension comes from the charged fermion masses. For model~A, the $\chi^2$ value is dominated by the pull of $m_\mu$, whereas for model~B, the largest pulls originate in the masses of the heaviest generation of charged fermions. Specifically, there seems to be some tension in that $m_\tau$ is predicted to be too large, while $m_t$ and $m_b$ are predicted to be too small. 

\begin{table}
\setlength{\tabcolsep}{1.5em}
    \begin{center}
    \begin{tabular}{ c @{}S@{} @{}S@{} @{}S@{} @{}S@{} }
    \hline \hline
         Observable & \multicolumn{2}{c}{Model A} & \multicolumn{2}{c}{Model B} \\ 
         							& {Value} & Pull & {Value} & Pull \\ \hline
$m_u$ [MeV]               & 1.22  &-0.0250& 1.27 & 0.188 \\
$m_c$ [GeV]                    & 0.620  &-0.0149& 0.657 & 1.20  \\
$m_t$  [GeV]                    & 164  &-0.481 & 158 & -1.21 \\
$m_d$ [MeV]                     & 2.64 &-0.171& 2.48 & -0.978\\
$m_s$ [MeV]                     & 53.4  &0.0519 & 46.5 & -1.46\\
$m_b$ [GeV]                     & 2.84  &-0.0177 & 2.49 & -2.53\\
$m_e$ [MeV]                     & 0.484  &0.0211 & 0.483 & 0.0125 \\
$m_\mu$ [GeV]                   & 0.0152  &-1.70& 0.0539 & -0.943 \\
$m_\tau$ [GeV]                  & 1.73  &0.0212 & 2.00 & 3.14 \\
$\Delta m^2_{21}$ [$10^{-5}\ \text{eV}^2$]  & 7.44  &0.0462 & 7.38 & -0.113\\
$\Delta m^2_{31}$ [$10^{-5}\ \text{eV}^2$]  & 2.52  &0.0385 & 2.55 & 0.271 \\
$|V_{12}|$                   & 0.226  &0.0535 & 0.224 & -0.0620\\
$|V_{13}|$                   & 0.00368  &0.0202 & 0.00379 & 0.613 \\
$|V_{23}|$                   & 0.0417  &0.0524 & 0.0396 & -0.967\\
$J$ [$10^{-5}$]                        & 3.06  &-0.0307& 3.07 & 0.0510 \\
$\sin^2\theta^\ell_{12}$   & 0.304  &0.00235 & 0.318 & 0.927  \\
$\sin^2\theta^\ell_{13}$   & 0.0221  &0.00307 & 0.0219 & -0.187\\
$\sin^2\theta^\ell_{23}$   & 0.569  &-0.0225& 0.586 & 0.588 \\
$\lambda$                  & 0.557  &-0.0281& 0.555 & -0.107\\ \hline
		$\chi^2$                       &  &3.17   & & 25.8  \\ \hline\hline
    \end{tabular}
    \end{center}
    \caption{\label{tab:pulls}Pulls of each observable included in the fit, for models A and B with top-down RG running. Defined such that a positive pull implies that the predicted value is larger than the corresponding true value.}
\end{table}

We also compute the predicted values of the masses of the light neutrinos $m_i$, and the heavy right-handed neutrino $M_i$, the effective neutrinoless double beta decay mass parameter $\langle m_{\beta\beta}\rangle$, and the  leptonic Dirac CP-violating phase $\delta_\text{CP}$ (observable in neutrino oscillations). For model~A, they are $m_i = \{8.97\times10^{-3}, 1.24\times10^{-2}, 5.09\times 10^{-2}\}\,\text{eV}$, $M_i = \{3.28\times10^{10}, 6.99\times10^{11}, 1.52\times10^{13}\}\,\text{GeV}$, $\langle m_{\beta\beta}\rangle = 3.60\times10^{-3}\,\text{eV}$, and $\delta_\text{CP}=0.117\pi$, whereas for model~B, they are $m_i = \{5.40\times10^{-3}, 1.01\times10^{-2}, 5.07\times 10^{-2}\}\,\text{eV}$, $M_i = \{7.94\times10^{8}, 4.89\times10^{10}, 1.11\times10^{13}\}\,\text{GeV}$, $\langle m_{\beta\beta}\rangle = 3.29\times10^{-3}\,\text{eV}$, and $\delta_\text{CP}=0.414\pi$. Notably, the sum of the masses of the light neutrinos are well within the limits set by cosmological observations~\cite{Aghanim:2018eyx} and the predicted value of $\langle m_{\beta\beta}\rangle$ is within current experimental bounds~\cite{Dolinski:2019nrj}. On the other hand, the predicted values of $\delta_\text{CP}$ in these two models deviate from the preferred values of global fits~\cite{Esteban:2020cvm}, as also found in Ref.~\cite{Ohlsson:2019sja}. However, one should keep in mind that these are not strict predictions of the models and that it is still possible that a fit that includes $\delta_\text{CP}$ is able to accommodate the favored value.

The parameter values that produce the fits in Tab.~\ref{tab:pulls} for model~A are 
\begin{mdframed}[align=center,userdefinedwidth=0.95\textwidth]
\begin{align*}
H &= 
\begin{psmallmatrix}
-1.64102\times10^{-6} - 4.07123\times10^{-6}\ii && -1.07109\times10{-4} - 1.29783\times10^{-4} \ii && 0 \\
-1.07109\times10{-4} - 1.29783\times10^{-4} \ii && -2.35762\times10^{-3} - 6.18487\times10^{-3}\ii && 0 \\
0 && 0 && -5.88890\times10^{-5} + 2.19376\times10^{-5}\ii
\end{psmallmatrix}\,,\\
F &=
\begin{pmatrix}
3.25857\times10^{-6} && 0 && 0 \\
0 && 1.53510\times10^{-3} && 0 \\
0 && 0 && -6.95026\times10^{-5} \\
\end{pmatrix}\,,\\
G &=
\begin{psmallmatrix}
0 && 0 && 2.78241\times10^{-6} - 2.03664\times10^{-6}\ii \\
0 && 0 && 1.39151\times10^{-4} - 3.40620\times10^{-20}\ii \\
-2.78241\times10^{-6} + 2.03664\times10^{-6}\ii && -1.39151\times10^{-4} + 3.40620\times10^{-20}\ii && 0 \\
\end{psmallmatrix}\,,
\end{align*}
\begin{alignat*}{4}
r &= 65.6885,   &&s= -0.783926 + 0.405357\mathrm{i},  &&&t_u = 2.80594-6.872567\times10^{-16}\ii,\\
v_d &= 3.86434\,\text{GeV}, && t_\ell = -8.30874-5.56461\ii,  &&&t_D = 2.96489-7.26189\times10^{-16}\ii, \\
 r_R &= 1.00584\times10^{16}\,\mathrm{GeV}, &&\quad M_\Delta = 7.17963\times10^{11}\,\mathrm{GeV},  &&&\quad  r_L = -4.17591\times10^{-9} - 1.65394\times10^{-10}\ii.
\end{alignat*}
\end{mdframed}
For model~B, the parameter values are
\begin{mdframed}[align=center,userdefinedwidth=0.95\textwidth]
\begin{align*}
H &= 
\begin{psmallmatrix}
0 && 0 && 2.64415\times10^{-6} + 1.33543\times10^{-6}\ii \\
0 && 0 && -1.09025\times10^{-3} - 5.34157\times 10^{-4}\ii \\
2.64415\times10^{-6} + 1.33543\times10^{-6}\ii && -1.09025\times10^{-3} - 5.34157\times 10^{-4}\ii && 0
\end{psmallmatrix}\,,\\
F &=
\begin{pmatrix}
-3.40218\times10^{-7} && 0 && 0 \\
0 && -5.15499\times10^{-3} && 0 \\
0 && 0 && 2.10311\times10^{-5} \\
\end{pmatrix}\,,\\
G &=
\begin{psmallmatrix}
0 && 0 && -2.66678\times10^{-7} - 1.38123\times10^{-6}\ii \\
0 && 0 && -8.75849\times10^{-4} - 4.34856\times10^{-4}\ii \\
2.66678\times10^{-7} + 1.38123\times10^{-6}\ii && 8.75849\times10^{-4} + 4.34856\times10^{-4}\ii && 0 \\
\end{psmallmatrix}\,,
\end{align*}
\begin{alignat*}{4}
r &= 7.94823,  && s = -6.57858 - 7.53804\mathrm{i},  &&&t_u = 4.03305-1.01362\times10^{-7}\ii,\\
v_d &= 0.172197\,\text{GeV}, && t_\ell = -0.955270+1.72986\ii, &&&t_D = 7.36091+23.4157\ii,\\
r_R &= 2.33462\times10^{15}\,\mathrm{GeV}, && \quad M_\Delta = 5.96059\times10^{13}\,\mathrm{GeV}, &&&\quad r_L = 8.43906\times10^{-10}+5.36491\times10^{-10}\ii.
\end{alignat*}
\end{mdframed}
In both of these models, the dominant contribution to the neutrino masses stems from the type I seesaw mechanism. By comparing the maximum singular values of the contributions in Eqs.~\eqref{eq:tI} and \eqref{eq:tII} evaluated from the parameters at $\MGUT$, we observe that the type I seesaw mechanism dominates by a factor of about 400 in model~A and a factor of about 13 in model~B. These numbers may, of course, change due to RG running, but the general conclusion should still hold.

\section{Summary and Conclusions}\label{sec:conclusion}

In this work, we have analyzed a class of non-supersymmetric $\SO10$ models with flavor symmetries, following the framework described in Ref.~\cite{Ferreira:2015jpa}. In the numerical fits, we have solved the RGEs for the sampled parameters from $\MGUT$ down to  $\MZ$ in order to fit to the known parameters at $\MZ$. This approach allows one to properly integrate out the heavy right-handed neutrinos and the scalar triplet, thereby taking into account the RG running of the neutrino parameters in an effective field theory framework, as opposed to approximate approaches in which the known parameters are extrapolated up to $\MGUT$. 

Out of the 14 models, only two have been found to accommodate the known values of fermion masses and mixing parameters. These are the ones labeled as model~A and model~B which both possess a $\mathbb{Z}_2$ symmetry, in agreement with previous results~\cite{Ferreira:2015jpa} that were found with supersymmetric models and bottom-up RG running. These two models are the ones with the largest number of free parameters, and are hence expected to provide the best fits. The difference between Ref.~\cite{Ferreira:2015jpa} and this work is that we considered non-supersymmetric models, that we used updated data, and most importantly, that we solved the RGEs of each sampled set of parameters from $\MGUT$ down to $\MZ$.

We have only considered the symmetries of the Yukawa potential in this work, and not the effect of the symmetries on the scalar potential. However, imposing a $\mathbb{Z}_2$ symmetry on the scalar potential should not present any major difficulties, and will, in general, mean that the $\rep{45}$ representation involved in the symmetry breaking (not analyzed in detail here) will also inherit this symmetry. It remains to construct a complete model with these flavor symmetries that accounts for symmetry breaking and gauge coupling unification. Finally, it would be interesting to investigate whether the disagreement between the predicted value of the CP-violating phase and that from global fits is a true prediction of the models or if it is possible to accommodate it by including it as one of the parameters in the fit.

\begin{acknowledgments}
We would like to thank Lu{\'i}s Lavoura for useful discussions and Mattias Blennow for supplying us with computing resources. T.O.~acknowledges support by the Swedish Research Council (Vetenskapsrådet) through contract No.~2017-03934 and the KTH Royal Institute of Technology for a sabbatical period at the University of Iceland. Numerical computations were performed on resources provided by the Swedish National Infrastructure for Computing (SNIC) at PDC Center for High Performance Computing (PDC-HPC) at KTH Royal Institute of Technology in Stockholm, Sweden under project numbers SNIC 2020/5-416 and SNIC 2021/5-141.
\end{acknowledgments}

\appendix
\section{Yukawa Textures}\label{sec:app_models}
In this appendix, we list the Yukawa matrix textures corresponding to the 14 models in Tab.~\ref{tab:models} as Eqs.~\eqref{eq:modelA}--\eqref{eq:modelE}. A ``$\times$'' symbol signifies a general non-zero matrix element. For the symmetry transformations and groups corresponding to each model, see Ref.~\cite{Ferreira:2015jpa}.

\begin{equation}\label{eq:modelA}
\text{A}:\quad Y_{10} \sim \begin{pmatrix}
\times & \times & 0 \\
\times & \times & 0 \\
0 & 0 & \times
\end{pmatrix}, \quad
Y_{126} \sim \begin{pmatrix}
\times & \times & 0 \\
\times & \times & 0 \\
0 & 0 & \times
\end{pmatrix},\quad 
Y_{120}\sim \begin{pmatrix}
0 & 0 & \times \\
0 & 0 & \times \\
\times & \times & 0
\end{pmatrix},
\end{equation}

\begin{equation}
\text{A}_1:\quad Y_{10} \sim \begin{pmatrix}
\times & 0 & 0 \\
0 & \times & 0 \\
0 & 0 & 0
\end{pmatrix}, \quad
Y_{126} \sim \begin{pmatrix}
0 & \times & 0 \\
\times & 0 & 0 \\
0 & 0 & \times
\end{pmatrix},\quad 
Y_{120}\sim \begin{pmatrix}
0 & 0 & \times \\
0 & 0 & 0 \\
\times & 0 & 0
\end{pmatrix}, 
\end{equation}

\begin{equation}
\text{A}_1':\quad Y_{10} \sim \begin{pmatrix}
\times & 0 & 0 \\
0 & 0 & 0 \\
0 & 0 & 0
\end{pmatrix}, \quad
Y_{126} \sim \begin{pmatrix}
0 & \times & 0 \\
\times & 0 & 0 \\
0 & 0 & \times
\end{pmatrix},\quad 
Y_{120}\sim \begin{pmatrix}
0 & 0 & \times \\
0 & 0 & 0 \\
\times & 0 & 0
\end{pmatrix}, 
\end{equation}

\begin{equation}
\text{A}_1'':\quad Y_{10} \sim \begin{pmatrix}
0 & 0 & 0 \\
0 & \times & 0 \\
0 & 0 & 0
\end{pmatrix}, \quad
Y_{126} \sim \begin{pmatrix}
0 & \times & 0 \\
\times & 0 & 0 \\
0 & 0 & \times
\end{pmatrix},\quad 
Y_{120}\sim \begin{pmatrix}
0 & 0 & \times \\
0 & 0 & 0 \\
\times & 0 & 0
\end{pmatrix}, 
\end{equation}

\begin{equation}
\text{A}_2:\quad Y_{10} \sim \begin{pmatrix}
0 & \times & 0 \\
\times & 0 & 0 \\
0 & 0 & \times
\end{pmatrix}, \quad
Y_{126} \sim \begin{pmatrix}
0 & \times & 0 \\
\times & 0 & 0 \\
0 & 0 & \times
\end{pmatrix},\quad 
Y_{120}\sim \begin{pmatrix}
0 & 0 & \times \\
0 & 0 & 0 \\
\times & 0 & 0
\end{pmatrix},
\end{equation}

\begin{equation}
\text{B}:\quad Y_{10} \sim \begin{pmatrix}
0 & 0 & \times\\
0 & 0 & \times \\
\times & \times & 0
\end{pmatrix}, \quad
Y_{126} \sim \begin{pmatrix}
\times & \times & 0\\
\times & \times & 0\\
0 & 0 & \times
\end{pmatrix},\quad 
Y_{120}\sim \begin{pmatrix}
0 & 0 & \times \\
0 & 0 & \times \\
\times & \times & 0
\end{pmatrix}, 
\end{equation}

\begin{equation}
\text{C}:\quad Y_{10} \sim \begin{pmatrix}
0 & \times & 0 \\
\times & 0 & \times \\
0 & \times & 0
\end{pmatrix}, \quad
Y_{126} \sim \begin{pmatrix}
\times & 0 & \times \\
0 & \times & 0 \\
\times & 0 & \times
\end{pmatrix},\quad 
Y_{120}\sim \begin{pmatrix}
0 & 0 & \times \\
0 & 0 & 0 \\
\times & 0 & 0
\end{pmatrix},
\end{equation}

\begin{equation}
\text{D}_1:\quad Y_{10} \sim \begin{pmatrix}
0 & \times & 0 \\
\times & 0 & 0 \\
0 & 0 & \times
\end{pmatrix}, \quad
Y_{126} \sim \begin{pmatrix}
0 & 0 & \times \\
0 & \times & 0 \\
\times & 0 & 0
\end{pmatrix},\quad 
Y_{120}\sim \begin{pmatrix}
0 & 0 & \times \\
0 & 0 & 0 \\
\times & 0 & 0
\end{pmatrix}, 
\end{equation}

\begin{equation}
\text{D}_2:\quad Y_{10} \sim \begin{pmatrix}
0 & \times & 0 \\
\times & 0 & 0 \\
0 & 0 & \times
\end{pmatrix}, \quad
Y_{126} \sim \begin{pmatrix}
\times & 0 & 0 \\
0 & 0 & \times \\
0 & \times & 0
\end{pmatrix},\quad 
Y_{120}\sim \begin{pmatrix}
0 & 0 & \times \\
0 & 0 & 0 \\
\times & 0 & 0
\end{pmatrix},
\end{equation}

\begin{equation}
\text{D}_3:\quad Y_{10} \sim \begin{pmatrix}
0 & 0 & \times  \\
0 & \times & 0 \\
\times & 0 & 0 
\end{pmatrix}, \quad
Y_{126} \sim \begin{pmatrix}
\times & 0 & 0 \\
0 & 0 & \times \\
0 & \times & 0
\end{pmatrix},\quad 
Y_{120}\sim \begin{pmatrix}
0 & 0 & \times \\
0 & 0 & 0 \\
\times & 0 & 0
\end{pmatrix},
\end{equation}

\begin{equation}
\text{D}_1':\quad Y_{10} \sim \begin{pmatrix}
0 & \times & 0 \\
\times & 0 & 0 \\
0 & 0 & 0
\end{pmatrix}, \quad
Y_{126} \sim \begin{pmatrix}
0 & 0 & \times \\
0 & \times & 0 \\
\times & 0 & 0
\end{pmatrix},\quad 
Y_{120}\sim \begin{pmatrix}
0 & 0 & \times \\
0 & 0 & 0 \\
\times & 0 & 0
\end{pmatrix}, 
\end{equation}

\begin{equation}
\text{D}_2':\quad Y_{10} \sim \begin{pmatrix}
0 & \times & 0 \\
\times & 0 & 0 \\
0 & 0 & 0
\end{pmatrix}, \quad
Y_{126} \sim \begin{pmatrix}
\times & 0 & 0 \\
0 & 0 & \times \\
0 & \times & 0
\end{pmatrix},\quad 
Y_{120}\sim \begin{pmatrix}
0 & 0 & \times \\
0 & 0 & 0 \\
\times & 0 & 0
\end{pmatrix},
\end{equation}

\begin{equation}
\text{D}_3':\quad Y_{10} \sim \begin{pmatrix}
0 & 0 & \times  \\
0 & 0 & 0 \\
\times & 0 & 0 
\end{pmatrix}, \quad
Y_{126} \sim \begin{pmatrix}
\times & 0 & 0 \\
0 & 0 & \times \\
0 & \times & 0
\end{pmatrix},\quad 
Y_{120}\sim \begin{pmatrix}
0 & 0 & \times \\
0 & 0 & 0 \\
\times & 0 & 0
\end{pmatrix},
\end{equation}

\begin{equation}\label{eq:modelE}
\text{E}:\quad Y_{10} \sim \begin{pmatrix}
0 & \times & 0 \\
\times & 0 & 0 \\
0 & 0 & 0
\end{pmatrix}, \quad
Y_{126} \sim \begin{pmatrix}
\times & 0 & 0 \\
0 & \times & 0 \\
0 & 0 & \times
\end{pmatrix},\quad 
Y_{120}\sim \begin{pmatrix}
0 & 0 & \times \\
0 & 0 & 0 \\
\times & 0 & 0
\end{pmatrix}.
\end{equation}

\bibliographystyle{apsrev4-1}
\bibliography{refs_flavor.bib}

\end{document}